# An Adaptive Reaction Force Observer Design

Emre Sariyildiz, *IEEE Student Member*, Kouhei Ohnishi, *IEEE Fellow*

***Abstract*— In this paper, a new adaptive design method is proposed for reaction force observer (RFOB) based robust force control systems. It is a well-known fact that a RFOB has several superiorities over a force sensor such as higher force control bandwidth, stability improvement, force-sensorless force control, and so on. However, there are insufficient analysis and design methods for a RFOB based robust force control system; therefore, its stability and performance highly depend on designers own experiences. To overcome this issue, a new stability analysis and a novel adaptive design methods are proposed for RFOB based robust force control systems. In the proposed adaptive design method, the design parameters of the robust force control system, i.e., the bandwidths of disturbance observer (DOB) and RFOB, the nominal and identified inertias in the design of DOB and RFOB, respectively, and the force control gain, are adjusted automatically by using an adaptive control algorithm which is derived by estimating the plant parameters and environmental impedance. The proposed adaptive design method provides good stability and performance by considering the design constraints of a DOB. The validity of the proposals is verified by simulation and experimental results.**

## I. INTRODUCTION

DEMANDS of dexterous and versatile mechanical systems, e.g., robots that can work with humans interactively, or robots that can make surgical operations, have been increased rapidly in the last decades [1, 2, 3 and 4]. It is a well-known fact that high performance motion control systems, i.e., position, force or admittance control systems, are one of the key points in the next generation robotics and mechatronics applications [5, 6]. Although position control problems have been solved successfully by using advanced robust control methods, e.g., sliding mode control, robustness and adaptability problems of force control systems are still challenging issues [5].

A Disturbance Observer (DOB) is a robust control tool that is widely used in motion control applications due to its simplicity and efficiency [7, 8 and 9]. A feed-back loop, namely inner-loop, is used to compensate system disturbances which are estimated by a DOB. To satisfy control goals, e.g., position or force control goals, an outer-loop controller can be designed by using the nominal plant parameters, since a DOB can nominalize the uncertain plant [7]. The bandwidth of a DOB and the ratios between uncertain and nominal plant parameters, e.g., inertia and torque coefficient, are the fundamental design parameters of a DOB [10, 11 and 12].

A reaction force observer (RFOB), which is used to estimate environmental impedance, is an application of a DOB [13 and 14]. It is designed by subtracting system uncertainties from the input of a DOB; therefore, a DOB and a RFOB have quite similar control structures. The main difference between a DOB and a RFOB is that the latter has a model based control structure, which is the most challenging issue in its design [13]. Superiorities of a RFOB over a force sensor, e.g., higher force control bandwidth, ideal-zero-stiffness force control, stability improvement, force-sensorless force control, etc., have been shown experimentally in the literature [13, 15 and 16]. Therefore, a RFOB is a quite effective motion control tool for the next generation robotics and mechatronics applications. However, its implementations suffer from insufficient analysis and design methods.

In this paper, a novel stability analysis and a new adaptive design methods are proposed for the RFOB based robust force control systems. The dynamics of a RFOB based robust force control system depend on the plant parameters, robustness and performance controllers and environmental impedance; therefore, they should be considered to achieve a high performance force control system. In the proposed stability analysis method, it is shown that a DOB and a RFOB can be designed as a phase lead-lag compensator, and the stability of the robust force control system can be improved by increasing the bandwidth of a RFOB. Besides that imperfect identification of a RFOB design is considered, and it is shown that not only the performance, but also the stability of the robust force control system changes significantly by the design parameters of a DOB and a RFOB. In the proposed adaptive design method, the design parameters of a RFOB based robust force control system are tuned automatically by using an online estimation algorithm of environmental impedance and plant parameters. It is shown that not only the force control gain, but also the bandwidths of a DOB and a RFOB, and the nominal and identified inertias in the design of a DOB and a RFOB, respectively, should be adjusted to improve the stability and performance. The proposed method provides good stability and performance for varying environmental impedance. The validity of the proposals is verified by simulation and experimental results.

The rest of the paper is organized as follows. In section II, a DOB and a RFOB are presented briefly. In section III, the stability of a RFOB based robust force control system is

Manuscript received September 20, 2013; revised January 12, 2014 and February 21, 2014; accepted April 7, 2014. This research was supported in part by the Ministry of Education, Culture, Sports, Science, and Technology of Japan under Grant- in-Aid for Scientific Research (S), 25220903, 2013.

E. Sariyildiz and K. Ohnishi are with the Ohnishi Laboratory, Department of System Design Engineering, Keio University, Yokohama, 223-8522, Japan. (e-mail:emre@sum.sd.keio.ac.jp, ohnishi@sd.keio.ac.jp)

analyzed. In section IV, a new adaptive RFOB design method is proposed. In section V, online parameter identification algorithms are proposed. In section VI, simulation and experimental results are given. The paper ends with conclusion given in the last section.

## II. Disturbance and Reaction Force Observers

### A. Disturbance Observer

A block diagram of a DOB based motion control system is shown in Fig.1. In this figure:

| | |
|---|---|
| $M_m, M_{mn}$ | Uncertain and nominal motor masses; |
| $K_F, K_{Fn}$ | Uncertain and nominal motor thrust coefficients; |
| $I_m, I_m^{des}, I_m^{cmp}$ | Total, desired and compensated motor currents; |
| $x_m, \dot{x}_m, \ddot{x}_m$ | Position, velocity and acceleration of motor; |
| $\ddot{x}_m^{des}$ | Desired motor acceleration; |
| $\dot{x}_m^{noise}$ | Noise of velocity measurement; |
| $g_{DOB}$ | Cut-off frequency of DOB; |
| $g_v$ | Cut-off frequency of velocity measurement; |
| $\Delta M_m$ | Motor mass variation; |
| $\Delta K_F$ | Motor thrust coefficient variation; |
| $F_m^{load}$ | Loading force; |
| $F_m^{frc}$ | Friction force; |
| $F_m^{int}$ | Interactive force; |
| $F_m^d$ | Total external disturbance; |
| $F_m^{dis}, \hat{F}_m^{dis}$ | Total system disturbance and its estimation; |

A DOB can estimate external disturbances and system uncertainties if they stay within its bandwidth $(g_{DOB})$ [7]. As shown in Fig.1, the estimated disturbances are fed-back so that the robustness of a motion control system is achieved.

The transfer functions of a DOB based motion control system are derived directly from Fig. 1 as follows:

If $g_v$ is infinite, i.e., perfect velocity measurement is achieved, then

$$\ddot{x}_m = \alpha \frac{s+g_{DOB}}{s+\alpha g_{DOB}} \ddot{x}_m^{des} - \frac{1}{M_m} T_{Sen}(s) F_m^d + T_{CoSen} s \dot{x}_m^{noise} \quad (1)$$

where $\alpha = \frac{M_{mn} K_F}{M_m K_{Fn}}$; $T_{Sen}(s) = \frac{1}{1+L_{DOB}(s)}$ and $T_{CoSen}(s) = \frac{L_{DOB}(s)}{1+L_{DOB}(s)}$ denote Sensitivity and Co-Sensitivity transfer functions, respectively; and $L_{DOB}(s) = \alpha \frac{g_{DOB}}{s}$. However, if $g_v$ is finite, then

$$\ddot{x}_m = \alpha \frac{(s+g_v)(s+g_{DOB})}{s^2 + g_v s + \alpha g_v g_{DOB}} \ddot{x}_m^{des} - \frac{1}{M_m} T_{Sen}(s) F_m^d + T_{CoSen} s \dot{x}_m^{noise} \quad (2)$$

where $T_{Sen}(s)$ and $T_{CoSen}(s)$ are same as defined above; however, $L_{DOB}(s) = \alpha \frac{g_v g_{DOB}}{s(s+g_v)}$.

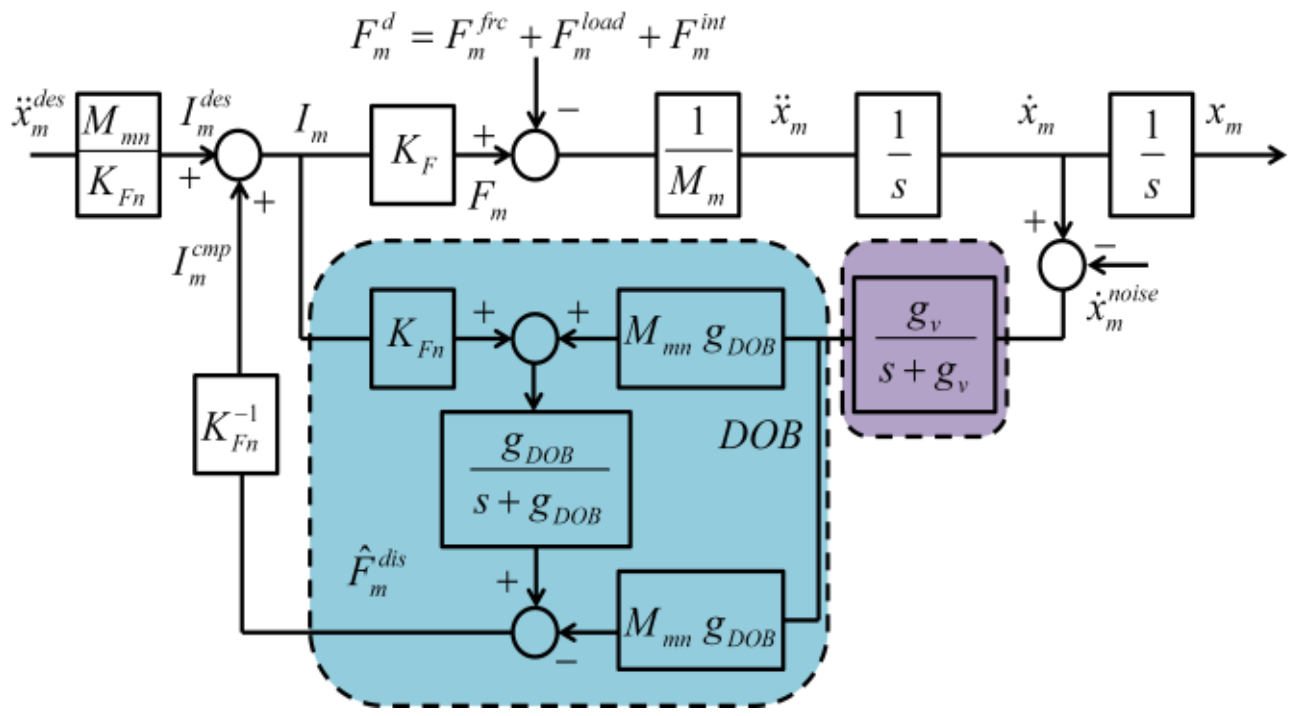

Fig. 1: A block diagram of a DOB

Equations (1) and (2) show that the derivative of the noise of velocity measurement gets transferred into the output by $T_{CoSen}$. Therefore, in general, the velocity measurement is filtered to suppress noise in the implementations of DOB based motion control systems [17]. Although it has never been considered, the low pass filter of velocity measurement changes the dynamics of the Sensitivity and Co-Sensitivity transfer functions at high frequencies, i.e., the robustness of a DOB, significantly.

The Bode integral theorem shows that if the relative degree of $L_{DOB}(s)$ is higher than one, then as the Sensitivity reduction at low frequencies is increased, the Sensitivity peak at high frequencies increases, i.e., a DOB becomes more sensitive to disturbances and noise at high frequencies [18 and 19]. Therefore, as shown in (2), $\alpha$ and $g_{DOB}$ cannot be increased freely when $g_v$ is finite [19 and 20]. New robustness bounds on $\alpha$ and $g_{DOB}$ can be proposed as follows:

Let us consider the Sensitivity and Co-Sensitivity transfer functions given in (2) by applying $g_v = \kappa g_{DOB}$.

$$T_{Sen} = \frac{s(s+\kappa g_{DOB})}{s^2 + \kappa g_{DOB} s + \alpha \kappa g_{DOB}^2}, \; T_{CoSen} = \frac{\alpha \kappa g_{DOB}^2}{s^2 + \kappa g_{DOB} s + \alpha \kappa g_{DOB}^2} \quad (3)$$

The characteristic function of (3) can be designed by using

$$w_n = \sqrt{\alpha \kappa}\, g_{DOB} \text{ and } \xi = 0.5\sqrt{\frac{\kappa}{\alpha}} \quad (4)$$

where $w_n$ and $\xi$ denote natural frequency and damping coefficient of a general second order characteristic polynomial, respectively. To bound the peaks of the Sensitivity and Co-Sensitivity transfer functions, if it is assumed that $\xi \geq 0.707$, then

$$\kappa \geq 2\alpha, \; or \; \alpha g_{DOB} \leq \frac{g_v}{2} \quad (5)$$

Equation (5) shows that $\alpha$ and $g_{DOB}$ are limited to bound the peaks of the Sensitivity and Co-Sensitivity transfer functions. The peak can be decreased by increasing $\xi$; however, the upper bounds of $\alpha$ and/or $g_{DOB}$ become more severe, and the performance and stability deteriorate [10 and 11]. Fig. 2 shows the constraints on the Sensitivity and Co-Sensitivity transfer functions. As shown in the figure, if $g_v$ is finite, then a DOB becomes more sensitive to disturbances and noise at high frequencies as $\alpha g_{DOB}$ is increased.

### B. Reaction Force Observer

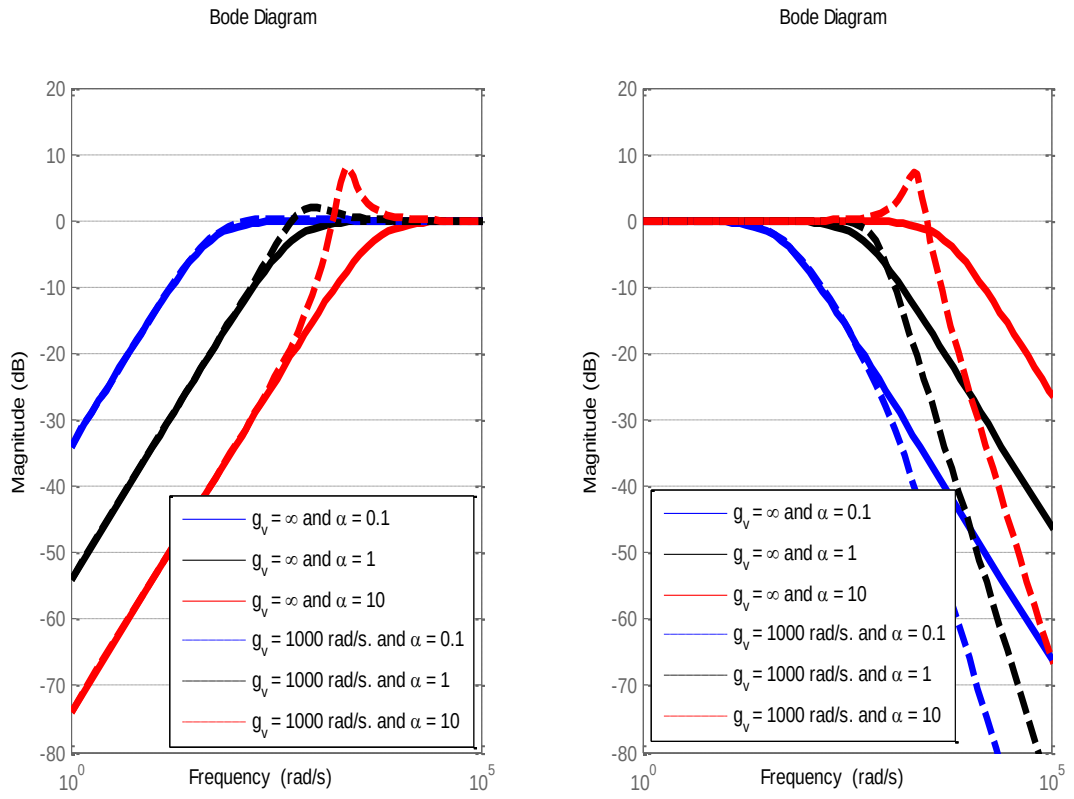

Fig. 2: Frequency responses of the Sensitivity and Co-Sensitivity functions

A RFOB, which is shown in Fig. 3., is used to estimate environmental impedance [13]. In this figure, $\hat{F}_m^{frc}$ and $\hat{F}_m^{int}$ denote the estimated friction and interactive forces, respectively; $\Delta\hat{M}_m$ and $\Delta\hat{K}_F$ denote the estimated mass and motor thrust coefficient variations, respectively; and $g_{RFOB}$ denotes the cut-off frequency of RFOB. The other parameters are same as defined above. As shown in Fig. 1 and Fig. 3, a DOB and a RFOB have quite similar control structures; however, only the latter requires the exact model of the uncertain plant [13].

## III. RFOB Based Robust Force Control System

A block diagram of a RFOB based robust force control system is shown in Fig. 4. In this figure, $C_f$ denotes a proportional force control gain, and the other parameters are same as defined above. In a RFOB based robust force control system, a DOB suppresses external disturbances and system uncertainties in the inner-loop so that a robust force control system is achieved. However, system uncertainties should be identified precisely in the design of a RFOB, i.e., in the outer-loop, to improve the stability and performance. The stability and performance of a RFOB based robust force control system can be analyzed as follows:

Environmental contact model is described by using a lumped spring-damper model as follows:

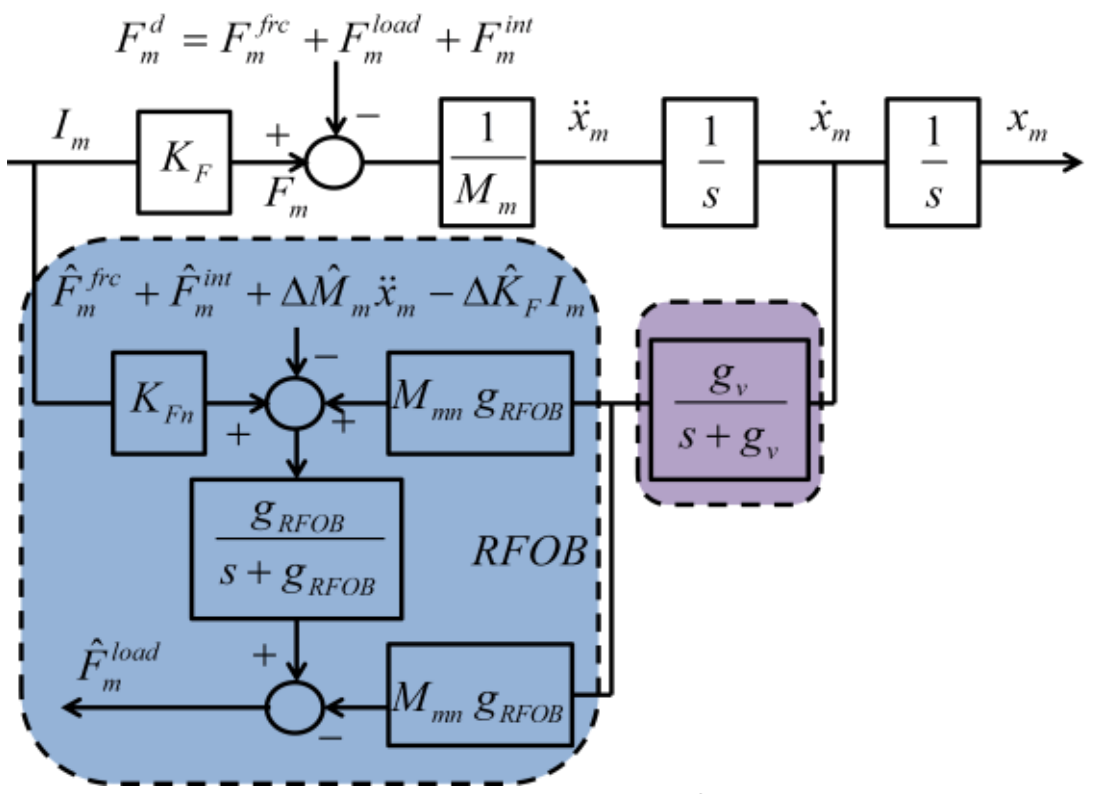

Fig. 3: A block diagram of a RFOB

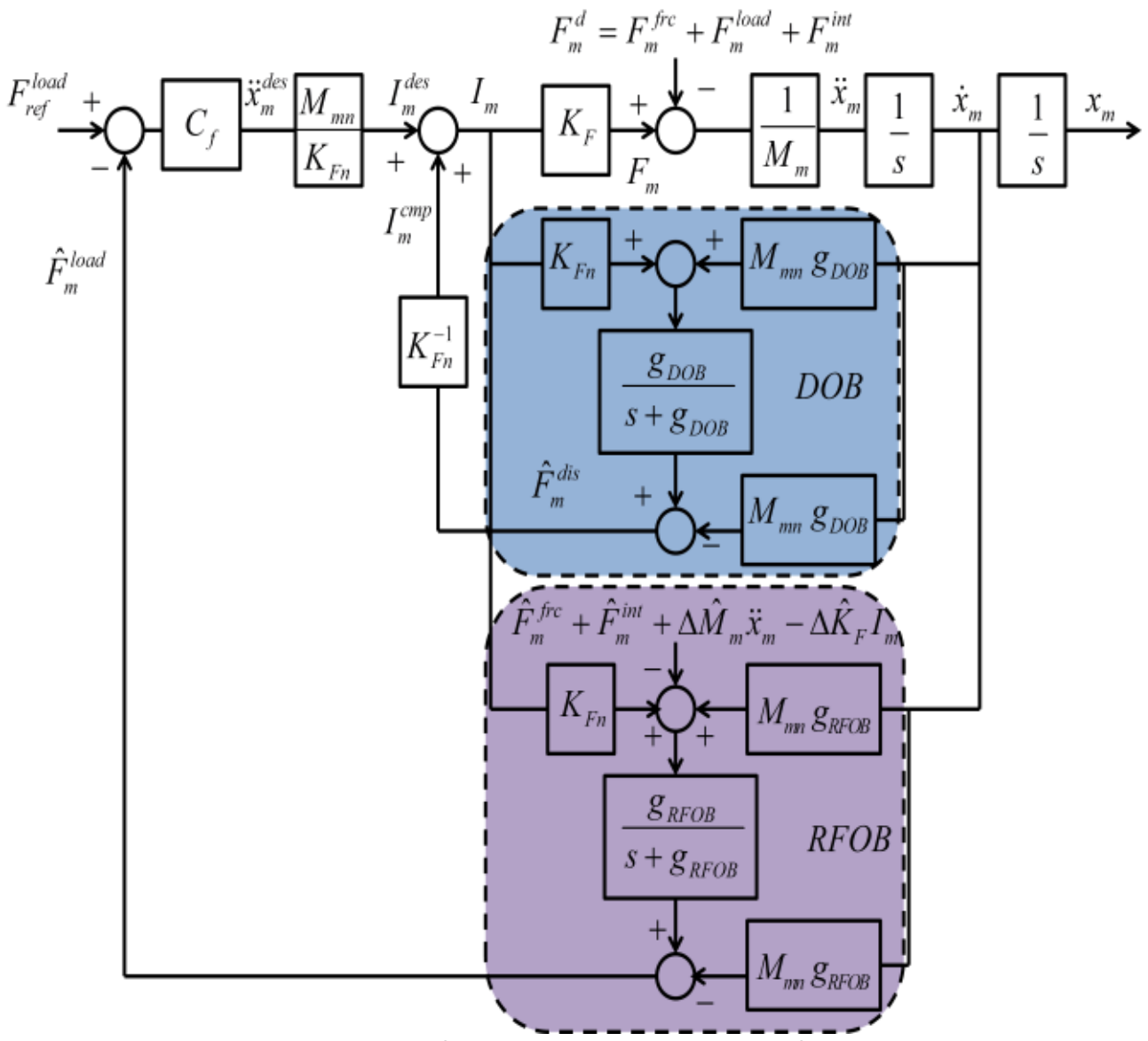

Fig. 4: A block diagram of a RFOB based robust force control system

$$F_m^{load} = D_{env}\left(\dot{x}_m - \dot{x}_{env}\right) + K_{env}\left(x_m - x_{env}\right) \quad (6)$$

where $D_{env}$ and $K_{env}$ denote the environmental damping and stiffness coefficients, respectively; and $x_{env}$ and $\dot{x}_{env}$ denote the position and velocity of environment at equilibrium, respectively. The transfer function between $F_{ref}^{load}$ and $\hat{F}_m^{load}$ is derived from Fig.4 as follows:

$$\frac{\hat{F}_m^{load}}{F_{ref}^{load}} = \frac{L_{RFOB}(s)}{1 + L_{RFOB}(s)} \quad (7)$$

where $L_{RFOB}(s) = C_f C_{cmp}(s) \dfrac{g_{RFOB}\dfrac{M_{mn}}{K_{Fn}}\varphi(s)}{s\left\{M_m s\left(s + \alpha g_{DOB}\right) + \left(D_{env}s + K_{env}\right)\right\}}$ (8)

denotes the open loop transfer function; $C_{cmp}(s) = \dfrac{\left(s + g_{DOB}\right)}{\left(s + g_{RFOB}\right)}$ is a phase lead-lag compensator; $\hat{K}_F = K_{Fn} + \Delta\hat{K}_F$ and $\hat{M}_m = M_{mn} + \Delta\hat{M}_m$ are the estimated thrust coefficient and inertia, respectively; and $\varphi(s) = \left(M_m\hat{K}_F - \hat{M}_m K_F\right)s^2 + \hat{K}_F D_{env}s + \hat{K}_F K_{env}$.

The relative degree of $L_{RFOB}(s)$ is one, so the asymptotes of the root loci are at angle of 180º. Let us define a new parameter by using $\beta = \dfrac{M_{mn}\hat{K}_F}{\hat{M}_m K_{Fn}}$. Equation (8) shows that if $\beta < \alpha$, then the open loop transfer function has a right half plane zero. Therefore, not only the performance, but also the stability of a RFOB based robust force control system may deteriorate by the imperfect system identification [11].

If a RFOB is designed by using a perfect system identification, i.e., $\alpha = \beta$, then the open-loop transfer function is

$$L_{RFOB}(s) = C_f C_{cmp}(s) \frac{g_{RFOB} M_m \alpha\left(D_{env}s + K_{env}\right)}{s\left\{M_m s\left(s + \alpha g_{DOB}\right) + \left(D_{env}s + K_{env}\right)\right\}} \quad (9)$$

The relative degree of $L_{RFOB}(s)$ is two, so the asymptotes of the root loci are at angle of ±90º . It is clear from (8) and (9) that the perfect system identification degrades the asymptotic behaves of the root loci.

The bandwidths of a DOB and a RFOB are generally set to the same value in the robust force control systems. If it is applied into (9), i.e., $g_{DOB} = g_{RFOB} = g$ , then

$$L_{RFOB}(s) = C_f \frac{gM_m\alpha\left(D_{env}s + K_{env}\right)}{s\left\{M_m s\left(s+\alpha g\right) + \left(D_{env}s + K_{env}\right)\right\}} \tag{10}$$

The relative degree of the open-loop transfer function is two, so the asymptotic behaves of the root loci do not change. However, the phase lead-lag compensator, $C_{cmp}(s)$, cannot be used to design the robust force control system.

Equations (8), (9) and (10) show that each of the open-loop transfer functions have a pole at the origin, so the steady state error is removed by a DOB in the robust force control systems.

Equation (8) shows that the stability of the robust force control system changes drastically by the imperfect identification of motor thrust coefficient and inertia. In practice, although thrust coefficient can be identified precisely, the identification of inertia may not be a simple task. Besides, the performance of the robust force control system is affected by the identification of thrust coefficient significantly, yet the error of inertia identification can be neglected due to small accelerations in many cases. Therefore, the paper proposes that a RFOB should be designed by using $\hat{M}_m \le M_m$ and $\hat{K}_F = K_F$ to improve the stability and performance, respectively.

Equation (7) shows that the robust force control system depends on the dynamics of the plant, environment, and robustness and performance controllers. Therefore, they should be considered in the design of a DOB based robust force control system.

## IV. Adaptive Reaction Force Observer Design

In this section, new adaptive design methods, which consider the practical design constraints of a DOB, will be proposed for RFOB based robust force control systems.

Let us start by considering damping environment.

### A. Damping Environment: $K_{env}$ is zero

If environmental impedance is considered as pure damping, then (10) and (7) are rewritten as follows:

$$L_{RFOB}(s) = C_f \frac{M_m\alpha g D_{env}}{M_m s^2 + \left(M_m\alpha g + D_{env}\right)s} \tag{11}$$

$$\frac{\hat{F}_m^{load}}{F_{ref}^{load}} = \frac{M_m C_f \alpha g D_{env}}{M_m s^2 + \left(M_m\alpha g + D_{env}\right)s + M_m C_f \alpha g D_{env}} \tag{12}$$

Let us consider a general second order transfer function model by using

$$C_{L_DES}(s) = \frac{w_n^2}{s^2 + 2\xi w_n s + w_n^2} \tag{13}$$

The design parameters of the robust force control system are derived as follows:

$$\alpha g = 2\xi w_n - \frac{D_{env}}{M_m} \tag{14}$$

$$C_f = \frac{w_n^2}{\alpha g D_{env}} \tag{15}$$

If the bandwidth constraint of a DOB, which is given in (5), is applied into (14), then

$$\frac{D_{env}}{M_m} < 2\xi w_n \le \frac{g_v}{2} + \frac{D_{env}}{M_m} \tag{16}$$

Consequently, the adaptive robust force control system is designed as follows:

- $\xi$ is chosen between $0.707 \le \xi \le 1$ to improve the stability and performance.
- $w_n$ is obtained by using $w_n = \frac{\gamma}{2\xi}\left(\frac{g_v}{2} + \frac{D_{env}}{M_m}\right)$ where $\frac{2D_{env}}{M_m g_v + 2D_{env}} < \gamma \le 1$ to satisfy (16).
- $\alpha g$ and $C_f$ are obtained by using (14) and (15).

Let us now consider stiff environment.

### B. Stiff Environment: $D_{env}$ is zero

If environmental impedance is considered as pure stiffness, then (10) and (7) are rewritten as follows:

$$L_{RFOB}(s) = C_f \frac{M_m\alpha g K_{env}}{s\left(M_m s^2 + M_m\alpha g s + K_{env}\right)} \tag{17}$$

$$\frac{\hat{F}_m^{load}}{F_{ref}^{load}} = C_f \frac{M_m\alpha g K_{env}}{M_m s^3 + M_m\alpha g s^2 + K_{env}s + M_m\alpha g C_f K_{env}} \tag{18}$$

Let us consider a desired characteristic polynomial by using

$$\begin{aligned} P_{CHDES}(s) &= (s+p)\left(s^2 + 2\xi w_n s + w_n^2\right) \\ &= s^3 + \left(2\xi w_n + p\right)s^2 + \left(w_n^2 + 2\xi w_n p\right)s + w_n^2 p \end{aligned} \tag{19}$$

The design parameters of the robust force control system are derived as follows:

$$p = \frac{K_{env} - M_m w_n^2}{2M_m\xi w_n} \tag{20}$$

$$\alpha g = 2\xi w_n + p \tag{21}$$

$$C_f = \frac{w_n^2 p}{\alpha g K_{env}} \tag{22}$$

If (5) is applied into (21), then

$$0 < 2\xi w_n + p \le \frac{g_v}{2} \tag{23}$$

Let us assume that $w_n = k\sqrt{\frac{K_{env}}{M_m}}$ where $k < 1$ to satisfy the stability. Then, (20) and (23) are rewritten as follows:

$$p = \eta\xi w_n = \frac{1-k^2}{2\xi k^2} w_n \quad (24)$$

$$w_n \le \frac{2\xi k^2}{1+\left(4\xi^2-1\right)k^2}\frac{g_v}{2} \quad (25)$$

where $\eta = \frac{p}{\xi w_n}$. $\eta$ is an important design parameter to adjust the performance of the system. If $k$ is derived in terms of $\eta$ by using (24) and is put into (25), then

$$k = \frac{1}{\sqrt{1+2\eta\xi^2}} \quad (26)$$

$$\eta^2 + \left(4-2R\right)\eta + 4 - \frac{R}{\xi^2} \le 0 \quad (27)$$

where $R = \frac{M_m g_v^2}{4K_{env}}$. The real and positive values of $\eta$ are derived from (27) if the following conditions are held.

**i.** If $\frac{M_m g_v^2}{K_{env}} \ge 16$, then $\xi$ can take any value.

**ii.** If $\frac{M_m g_v^2}{K_{env}} < 16$, then $\xi$ should satisfy $\xi \le \xi^*$ where $\xi^* = \frac{2\sqrt{K_{env}}}{\sqrt{16K_{env} - M_m g_v^2}}$ to obtain real $\eta$ and $\xi^* = 0.5\sqrt{\frac{M_m g_v^2}{4K_{env}}}$ to obtain $\eta > 0$.

Consequently, the adaptive robust force control system is designed as follows:

- $\eta$ is determined by considering **i** and **ii**. If $\xi^* < 1$, then $\eta$ should be chosen small enough, e.g. $\eta = 0.1$, to suppress the effects of low-damping poles; however, if $\xi^* > 1$, then $\eta$ can be chosen freely.
- $\xi$ is determined by using
  - If $4\left(2+\eta\right)^2 K_{env} \le 2M_m\eta g_v^2$, then $\xi$ can take any value
  - If $4\left(2+\eta\right)^2 K_{env} > 2M_m\eta g_v^2$, then

$$\xi^2 \le \frac{M_m g_v^2}{4K_{env}\left(2+\eta\right)^2 - 2\eta M_m g_v^2}$$

- $p$ and $k$ are obtained by using (24) and (26).
- $w_n$ is obtained by using $w_n = k\sqrt{\frac{K_{env}}{M_m}}$.
- $\alpha g$ and $C_f$ are obtained by using (21) and (22).

Lastly, let us consider stiff and damping environment.

*C. Stiff and Damping Environment:*

If environmental impedance is modeled by using damping and stiffness, then (7) is rewritten as follows:

$$\frac{\hat{F}_m^{load}}{F_{ref}^{load}} = \frac{C_f\alpha g\left(D_{env}s + K_{env}\right)}{M_m s^3 + \left(M_m\alpha g + D_{env}\right)s^2 + \left(C_f M_m\alpha g D_{env} + K_{env}\right)s + C_f M_m\alpha g K_{env}} \quad (28)$$

If (19) is considered, then the design parameters of the robust force control system are derived as follows:

$$p = \frac{K_{env}^2 - w_n^2 M_m K_{env}}{2\xi M_m w_n K_{env} - w_n^2 M_m D_{env}} \quad (29)$$

$$\alpha g = 2\xi w_n + p - \frac{D_{env}}{M_m} \quad (30)$$

$$C_f = \frac{w_n^2 p}{\alpha g K_{env}} \quad (31)$$

If (5) is applied into (30), then

$$\frac{D_{env}}{M_m} < 2\xi w_n + p \le \frac{g_v}{2} + \frac{D_{env}}{M_m} \quad (32)$$

Let us assume that $w_n = k\sqrt{\frac{K_{env}}{M_m}} = k^* 2\xi\frac{K_{env}}{D_{env}}$. Then, (29) is rewritten as follows:

$$p = \eta\xi w_n = \frac{1-k^2}{2\xi^2\left(1-\psi k\right)k^2}\xi w_n \quad (33)$$

where $\psi = \frac{\sqrt{\frac{K_{env}}{M_m}}}{2\xi\frac{K_{env}}{D_{env}}}$; and $\eta = \frac{p}{\xi w_n}$.

The stability of the robust force control system is achieved if

$$k < 1 \textit{ and } k < \frac{1}{\psi}, \textit{ or } k > 1 \textit{ and } k > \frac{1}{\psi} \quad (34)$$

and the bandwidth constraint of a DOB is satisfied if

$$\frac{D_{env}}{M_m} < \left(2+\eta\right)\xi k\sqrt{\frac{K_{env}}{M_m}} \le \frac{g_v}{2} + \frac{D_{env}}{M_m} \quad (35)$$

Against the pure damping and pure stiffness cases, it is not an easy task to design an adaptive RFOB analytically when environmental impedance is modeled by using damping and stiffness. To overcome this issue, a simple and effective design method is proposed as follows:

Let us consider the relation between $k$ and $\eta$ by using (33)

$$\eta = \frac{1-k^2}{2\xi^2\left(1-\psi k\right)k^2} \quad (36)$$

Equation (36) shows that $\eta$ is zero and infinite when $k$ is equal to one and $\frac{1}{\psi}$, respectively. Therefore, $k$ should be chosen close to one when $\eta$ is desired to be small enough to suppress low damping poles' effects. If it is considered, then the constraints on $\xi$ can be defined approximately as follows:

$$\xi^- < \xi \leq \xi^+ \quad (37)$$

where $\xi^- = \dfrac{\frac{D_{env}}{M_m}}{2\sqrt{\frac{K_{env}}{M_m}}}$ and $\xi^+ = \dfrac{\frac{g_v}{2} + \frac{D_{env}}{M_m}}{2\sqrt{\frac{K_{env}}{M_m}}}$.

If $\xi^+ < 1$, then the DOB constraints, i.e., $\xi^- < \xi \leq \xi^+$, and $\eta < 1$ should be satisfied; however, if $\xi^+ \geq 1$, then only the DOB constraints should be satisfied, i.e., there is no a constraint on $\eta$. Therefore, two different solutions should be considered to design an adaptive RFOB.

Consequently, the adaptive robust force control system is designed as follows:

- The constraints on $\xi$ are determined by using (37).
  - If $\xi^- < \xi \leq \xi^+$ where, e.g., $\xi^+ < 1$,
    - Chose $\xi$ in the given interval, and $\eta = \eta^*$ where $\eta^* < 1$.
    - Solve k by using (37)

$$2\eta^*\xi^2\psi k^3 - \left(1 + 2\eta^*\xi^2\right)k^2 + 1 = 0 \quad (38)$$

    - Chose the solution of $k$ which is close to 1.
  - If $\xi^- < \xi \leq \xi^+$ where, e.g., $\xi^+ \geq 1$,
    - Chose $\xi = 1$.
    - Solve k by using

$$\begin{aligned} &4\xi^2\psi k^3 + \left(1 - 4\xi^2 - 2\xi\psi\vartheta\right)k^2 + 2\xi\vartheta k - 1 \geq 0 \\ &4\xi^2\psi k^3 + \left(1 - 4\xi^2 - 2\xi\psi\delta\right)k^2 + 2\xi\delta k - 1 < 0 \end{aligned} \quad (39)$$

where $\vartheta = \dfrac{\frac{g_v}{2} + \frac{D_{env}}{M_m}}{\sqrt{\frac{K_{env}}{M_m}}}$ and $\delta = \dfrac{\frac{D_{env}}{M_m}}{\sqrt{\frac{K_{env}}{M_m}}}$

    - Chose the real and positive solutions of $k$.
- $w_n$ is obtained by using $w_n = k\sqrt{\dfrac{K_{env}}{M_m}}$.
- $p$, $\alpha g$ and $C_f$ are obtained by using (29-31).

The solutions of the cubic equations, which give real and positive $k$ values, are given in the Appendix.

## V. Parameter Identification

In this section, online parameter identification algorithms will be proposed.

The dynamic equation of a DOB based robust motion control system is written by using Fig. 1 as follows:

$$\left(\frac{M_{mn}}{K_{Fn}}\ddot{x}_m^{des} + \frac{\hat{F}_m^{dis}}{K_{Fn}}\right)K_F = M_m\ddot{x}_m + F_m^{frc} + F_m^{load} \quad (40)$$

where $F_m^{load} = D_{env}\left(\dot{x}_m - \dot{x}_{env}\right) + K_{env}\left(x_m - x_{env}\right)$. For the sake of simplicity, let us use the static model of friction; however, more complex models, such as Lugre friction model, can be implemented similarly [21]. The friction is modeled as follows:

$$F_m^{frc} = k_{vsc}\dot{x}_m + k_{clmb}\varsigma\left(\dot{x}_m\right) \quad (41)$$

where $k_{vsc}$ and $k_{clmb}$ denote the viscous and coulomb friction coefficients, respectively; $\varsigma\left(\dot{x}_m\right)$ denotes the approximation of the coulomb friction model [22].

To design an adaptive RFOB, the parameters to be determined are $M_m, k_{vsc}, k_{clmb}, D_{env}$ and $K_{env}$. Because external load is estimated by using a RFOB, the plant parameters and environmental impedance can be identified during non-contact and contact motions, separately, i.e., the plant parameters and environmental impedance cannot be identified, simultaneously.

Let us first consider the non-contact motion and identify the plant parameters. Equation (40) can be rewritten as follows:

$$u_{nc} = \boldsymbol{\rho}_{nc}^T\boldsymbol{\delta}_{nc} \quad (42)$$

where $u_{nc} = M_{mn}\ddot{x}_m^{des} + \hat{F}_m^{dis}$; $\boldsymbol{\rho}_{nc} = \left[\ddot{x}_m, \dot{x}_m, \varsigma\left(\dot{x}_m\right), 1\right]^T$; $\boldsymbol{\delta}_{nc} = \left[M_m, k_{vsc}, k_{clmb}, \hat{F}_m^d\right]^T$.

It is reasonable to assume that the unknown parameters are bounded by a convex set. Let us define the convex set by using $\forall\boldsymbol{\delta}_{nc}(i) \in \Xi_{nc}, \boldsymbol{\delta}_{nc}^{min}(i) \leq \boldsymbol{\delta}_{nc}(i) \leq \boldsymbol{\delta}_{nc}^{max}(i), i = 1,2,3,4$.

The recursive least mean square error algorithm (RLMS) is used to identify the plant parameters as follows:

$$\begin{aligned} &\mathbf{K}_{nc}(t) = \boldsymbol{\Gamma}_{nc}(t-1)\boldsymbol{\rho}_{nc}(t)\left(\mu_{nc} + \boldsymbol{\rho}_{nc}(t)^T\boldsymbol{\Gamma}_{nc}(t-1)\boldsymbol{\rho}_{nc}(t)\right)^{-1} \\ &\boldsymbol{\delta}_{nc}(t) = \boldsymbol{\delta}_{nc}(t-1) + Prj_{nc}\left\{\mathbf{K}_{nc}(t)\left(u_{nc}(t) - \boldsymbol{\rho}_{nc}(t)^T\boldsymbol{\delta}_{nc}(t-1)\right)\right\} \\ &\boldsymbol{\Gamma}_{nc}(t) = \frac{1}{\mu_{nc}}\left(\mathbf{I}_4 - \mathbf{K}_{nc}(t)\boldsymbol{\rho}_{nc}(t)^T\right)\boldsymbol{\Gamma}_{nc}(t-1) \end{aligned} \quad (43)$$

where $\mu_{nc}$ denotes forgetting factor; $\bullet_{nc}$ denotes the parameters in non-contact motion; and

$$Prj_{nc}\left\{\bullet_{nc}(i)\right\} = \begin{cases} 0, & \boldsymbol{\delta}_{nc}(i) \leq \boldsymbol{\delta}_{nc}^{min}(i) \\ 0, & \boldsymbol{\delta}_{nc}(i) \geq \boldsymbol{\delta}_{nc}^{max}(i) \\ \bullet_{nc}(i) & otherwise \end{cases} \quad (44)$$

The projection function $Prj_{nc}\left\{\boldsymbol{\delta}_{nc}(i)\right\}, i = 1, \cdots, 4$, provides that the plant parameters are updated only in non-contact motion and do not burst.

To estimate environmental impedance, (42) is rewritten similarly as follows:

$$u_c = \boldsymbol{\rho}_c^{\mathrm{T}}\boldsymbol{\delta}_c \quad (45)$$

where $u_c = \hat{F}_m^{load}$; $\boldsymbol{\rho}_c = \left[\dot{x}_m, x_m, 1\right]^T$; $\boldsymbol{\delta}_c = \left[D_{env}, K_{env}, \hat{F}_m^d\right]^T$. The environmental impedance is identified by using the RLMS as follows:

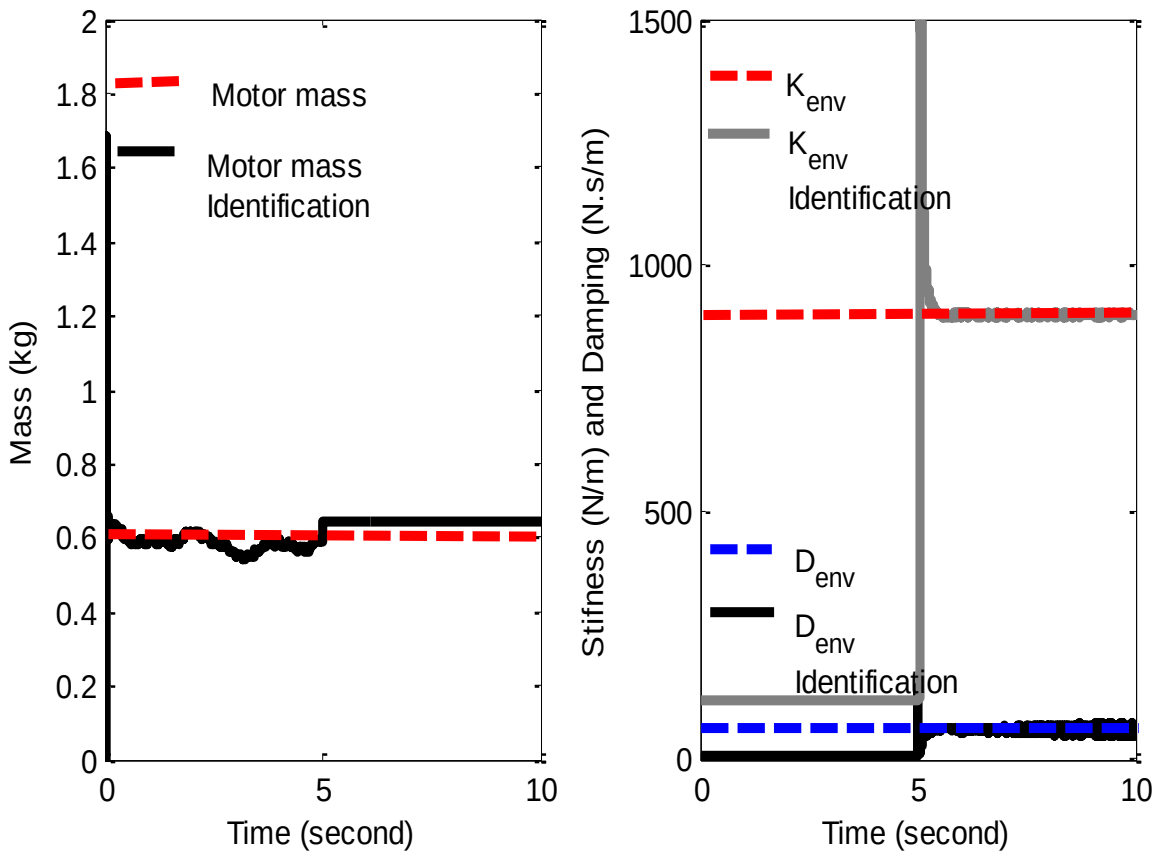

Fig. 5: Performance of the proposed identification algorithm

$$\begin{aligned}
\mathbf{K}_c(t) &= \mathbf{\Gamma}_c(t-1)\boldsymbol{\rho}_c(t)\left(\mu_c + \boldsymbol{\rho}_c(t)^T\mathbf{\Gamma}_c(t-1)\boldsymbol{\rho}_c(t)\right)^{-1} \\
\boldsymbol{\delta}_c(t) &= \boldsymbol{\delta}_c(t-1) + Prj_c\left\{\mathbf{K}_c(t)\left(u_c(t) - \boldsymbol{\rho}_c(t)^T\boldsymbol{\delta}_c(t-1)\right)\right\} \\
\mathbf{\Gamma}_c(t) &= \frac{1}{\mu_c}\left(\mathbf{I}_3 - \mathbf{K}_c(t)\boldsymbol{\rho}_c(t)^T\right)\mathbf{\Gamma}_c(t-1)
\end{aligned} \quad (46)$$

where $\mu_c$ denotes forgetting factor; $\bullet_c$ denotes the parameters in contact motion; and

$$Prj_c\{\bullet_c(i)\} = \begin{cases} 0, & \boldsymbol{\delta}_c(i) \le \boldsymbol{\delta}_c^{min}(i) \\ 0, & \boldsymbol{\delta}_c(i) \ge \boldsymbol{\delta}_c^{max}(i) \\ \bullet_c(i) & otherwise \end{cases} \quad (47)$$

It is obvious that the uncertainty range of environmental impedance is larger than the plant parameters' one. The projection function, $Prj_c\{\bullet_c(i)\}$, provides that the estimation of environmental impedance is conducted only in contact motion. In the proposed RLMS algorithm, the projection functions work discontinuously, and the parameters are updated conditionally.

Fig. 5 shows the performance of the proposed RLMS algorithm. During non-contact motion, the inertia of a linear motor is identified. To achieve contact motion, a known environmental impedance is designed by using zero position control of a dc motor, in which $K_p = 900$ and $K_d = 60$ are the parameters of the PD position controller. Fig. 5 indicates that the plant parameters and environmental impedance can be identified by using the proposed algorithm. It is obvious that the convergence rates of the parameters affect the performance of the adaptive RFOB. Besides, the impact force causes high identification errors initially in the environmental impedance identification. Therefore, the parameters of the adaptive RFOB should be updated by considering the drawbacks of the proposed on-line RLMS algorithm, i.e., the parameters should be updated when they converge. Fig. 6 shows the block diagram of the proposed adaptive RFOB based robust force control system.

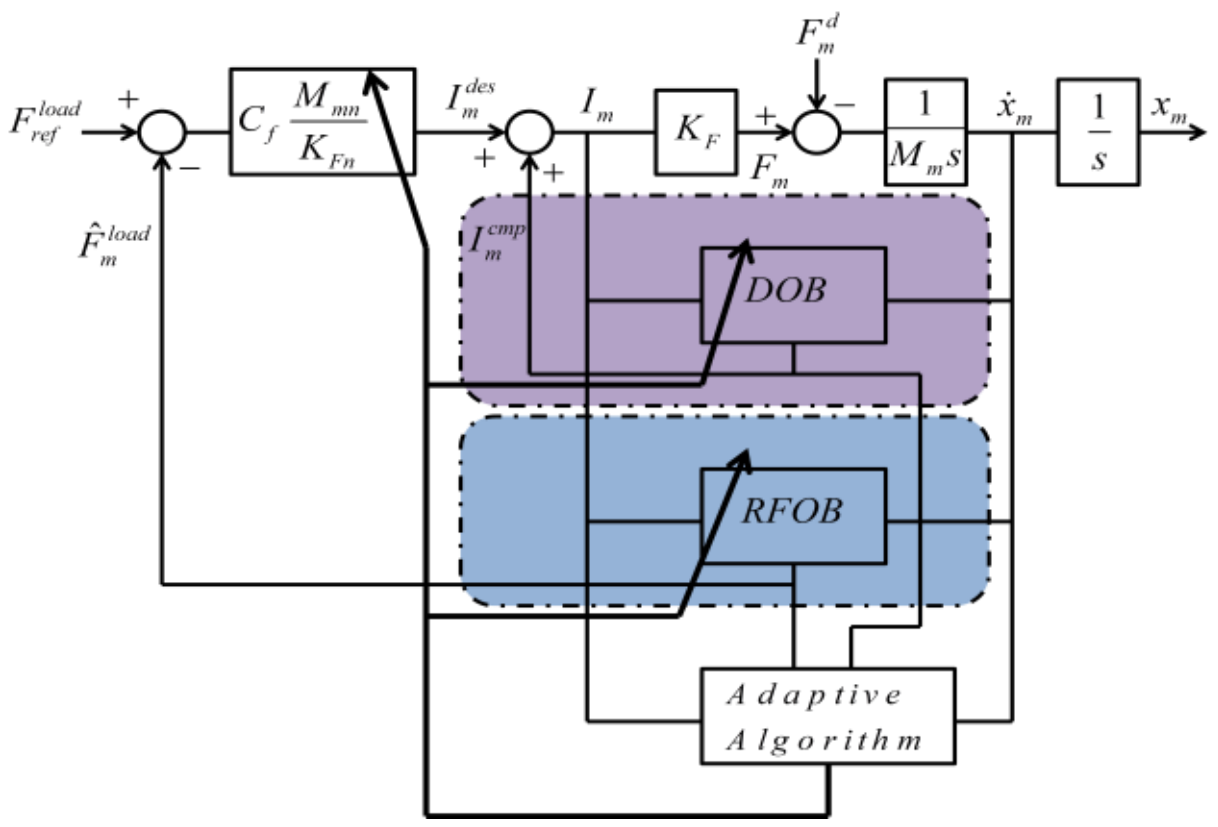

Fig. 6: Block diagram of the adaptive RFOB based force control system

TABLE I
SIMULATION PARAMETERS

| Parameters | Descriptions | Values |
|---|---|---|
| Mm | motor mass | 0.025 k.g. |
| $K_F$ | motor thrust coefficient | 0.5 N/A |
| $g_v$ | cut-off frequency of velocity measurement | 1000 rad/s. |

## VI. SIMULATION AND EXPERIMENT

In this section, simulation and experimental results will be presented.

### A. *Simulation*

The stability and performance of a RFOB based robust force control system are carried out in the simulations. The simulation parameters are shown in Table I.

Fig. 7 show the root-loci of a RFOB based robust force control system with respect to the force control gain $C_f$. Fig 7a and 7b are plotted by using (9) and (8), respectively, and it is

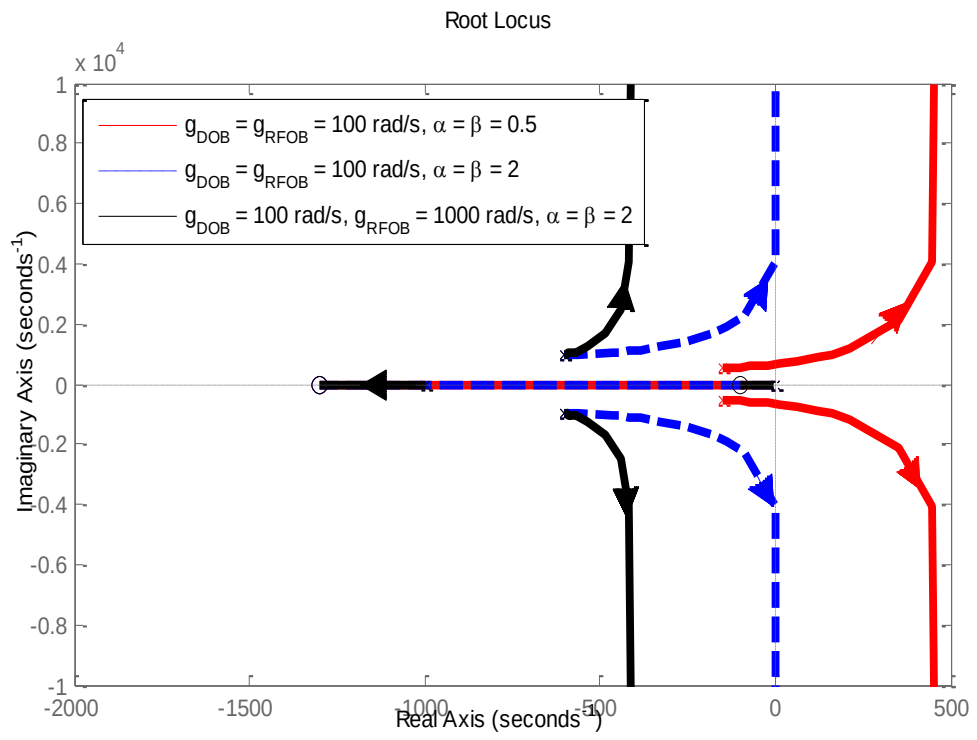

(a) $\alpha = \beta$

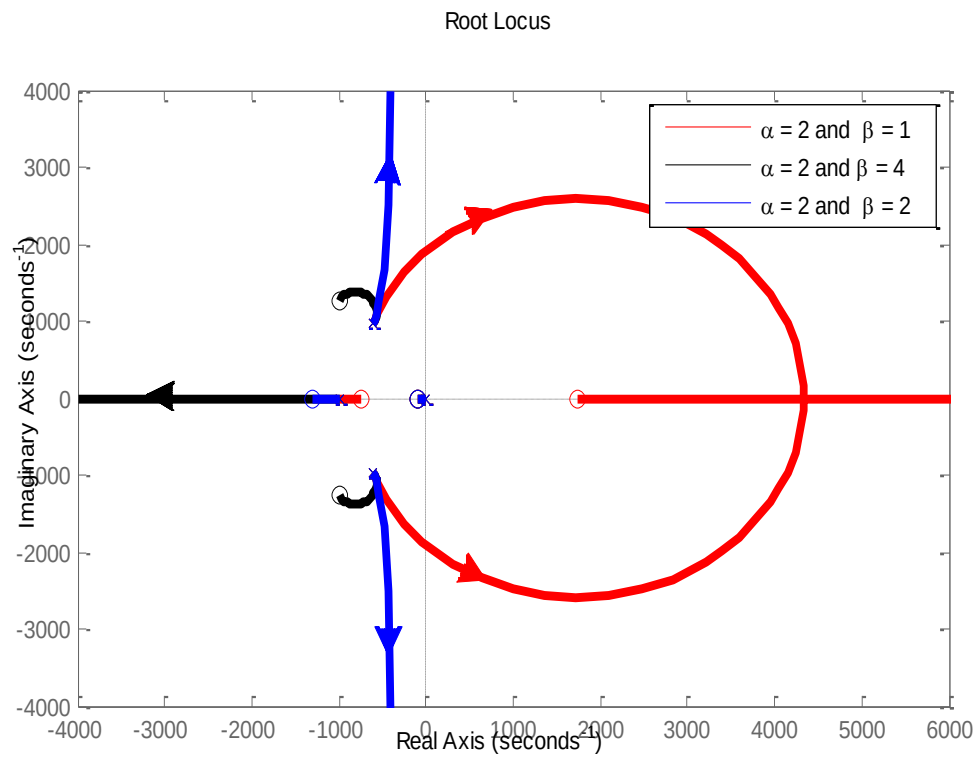

(b) $g_{DOB} = 100\,rad/s$ and $g_{RFOB} = 1000\,rad/s$

Fig. 7: Stability of a RFOB based robust force control system

assumed that $K_{env} = 6500\, N/m$ and $D_{env} = 2\, Ns/m$. Fig 7a indicates that the stability of the robust force control system can be improved by increasing $\alpha$ and $g_{RFOB}$. Fig. 7b indicates that the imperfect identification of inertia changes the stability of the robust force control system significantly. To improve the stability, $\alpha \le \beta$ should be guaranteed in the design of a RFOB.

Fig.8 shows the tunings of the design parameters by using the proposed adaptive algorithms. It is assumed that $\alpha = 1$, so the maximum bandwidth of a DOB is 500 rad/s. to achieve a good robustness. As shown in Fig. 8, the maximum bandwidth of a DOB can be achieved if damping environment is considered. However, if stiff environment is considered, then the bandwidth of a DOB should be limited to improve the stability and performance when the environmental stiffness is low. The performance and robustness of the force control system can be improved if the environmental stiffness is considered as damping and stiffness. In this case, the bandwidth of a DOB can be increased even if the environmental stiffness is low. Although there is a small pole near the origin, the performance of the force control system is not affected due to the zero near the pole.

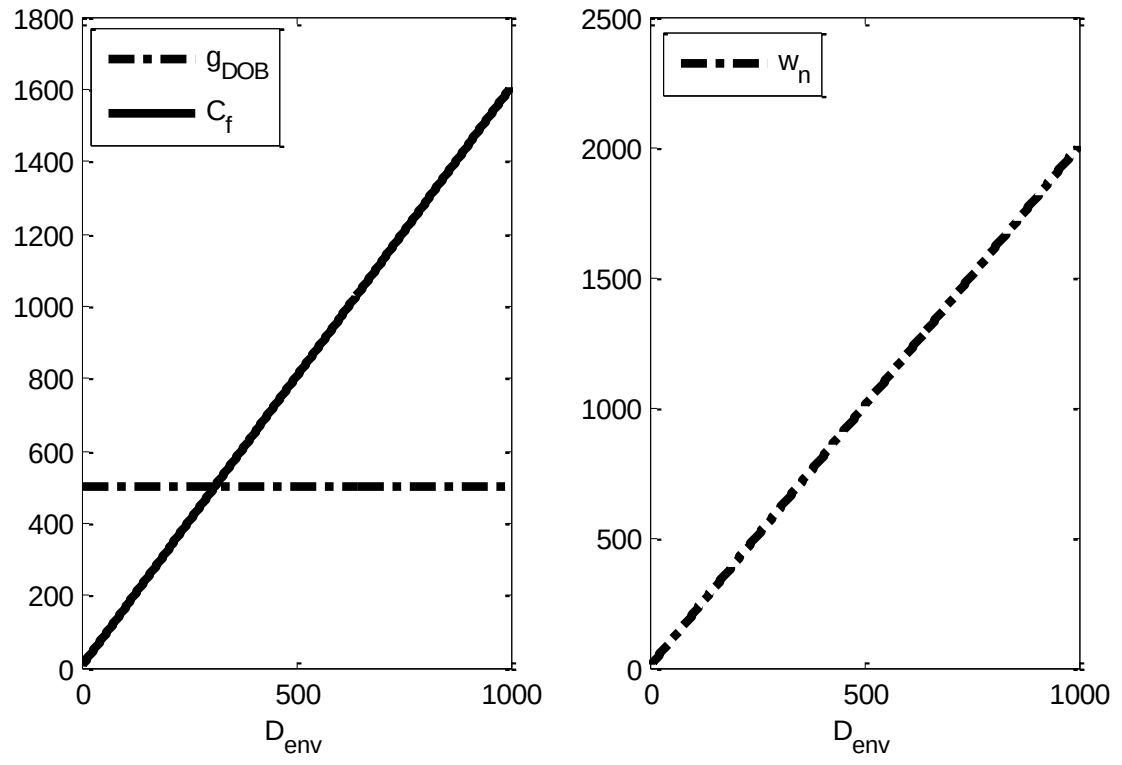

(a) $K_{env} = 0$ *and* $D_{env} \neq 0$

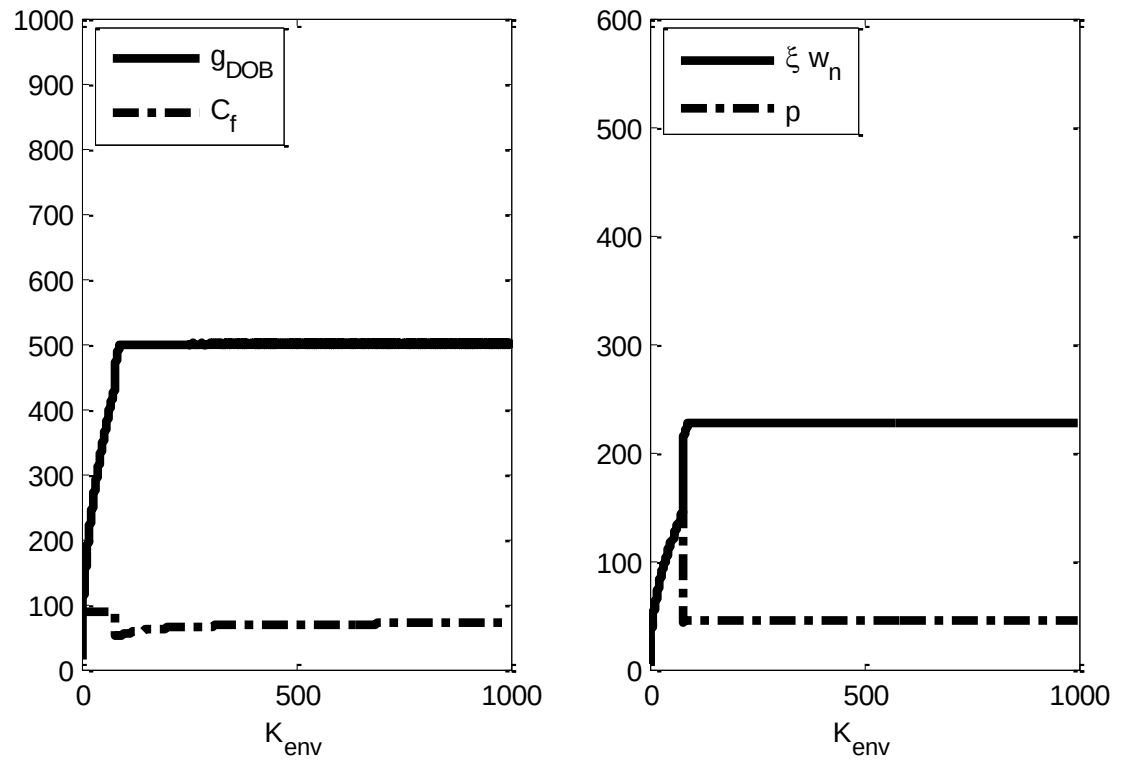

(b) $K_{env} \neq 0$ *and* $D_{env} = 0$

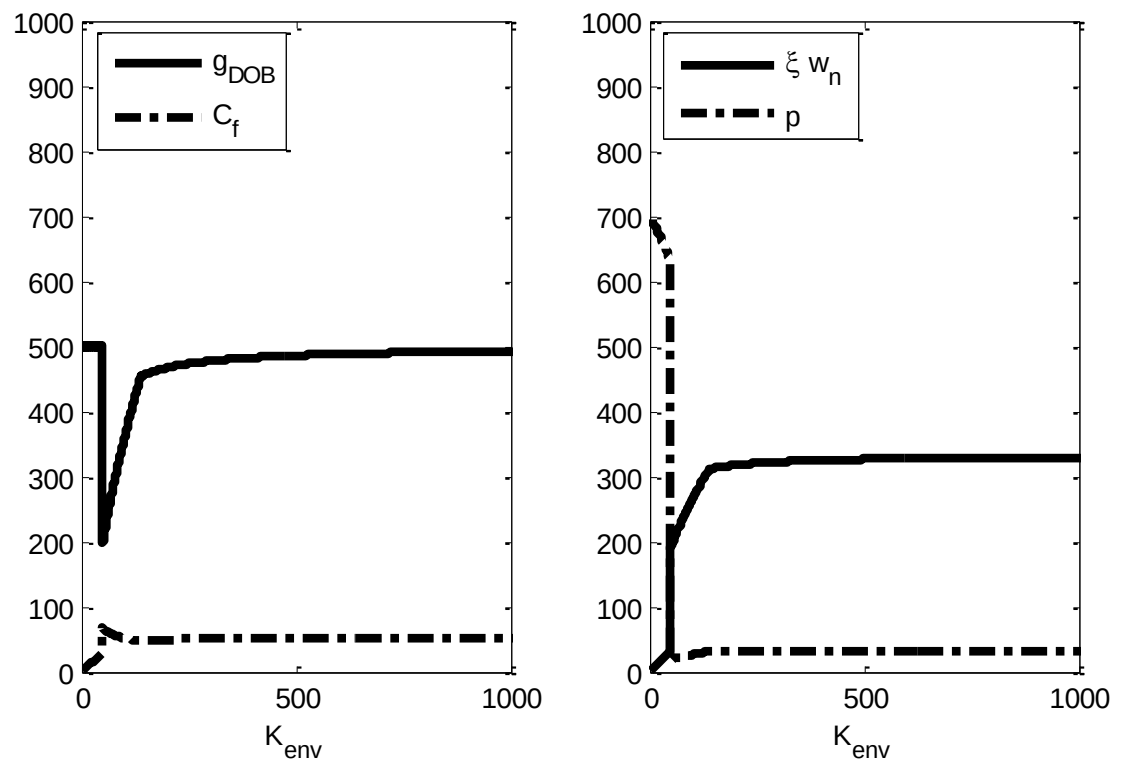

(c) $K_{env} \neq 0$ *and* $D_{env} = 0.25$

Fig. 8: Design parameters tuning and the poles of the force control system

TABLE II
EXPERIMENT PARAMETERS

| Parameters | Descriptions | Values |
|---|---|---|
| $M_{m1}$ | motor mass in the vertical direction | 0.81 k.g. |
| $M_{m2}$ | motor mass in the horizontal direction | 3.02 k.g. |
| $K_F$ | motor thrust coefficient | 0.5 N/A |
| $K_P$ | proportional gain of position control | 1200 |
| $K_V$ | derivative gain of position control | 90 |
| $g_v$ | cut-off frequency of velocity measurement | 1000 rad/s. |

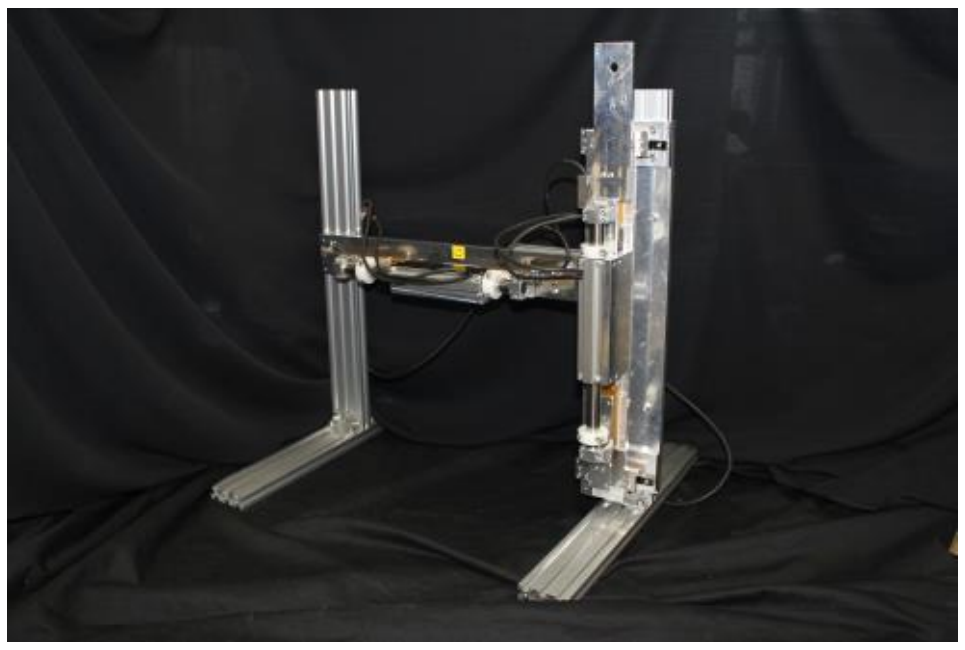

Fig. 9: A XZ-table mechanism

### B. Experiment

A XZ-table mechanism, which is shown in Fig.9, is carried out to show the validity of the proposals. The specifications of the experimental setup are shown in Table II. The sampling time is 0.1 ms. KYOWA LUR-A-50NSA1 force sensor is used to verify the performance of RFOB.

Let us start by considering how identification of plant parameters improves the performance of the robust force control system. In the vertical direction of table mechanism, force control is implemented between 0 to 5 and 10 to 15 seconds; position control is implemented between 5 to 10 seconds, and the uncertain plant parameters, i.e., motor mass and friction, are identified by considering gravity. Fig. 10 shows that the position and force control goals are achieved. The performance of the RFOB is improved between 10 to 15 seconds by identifying the plant parameters during non-contact motion. A soft environment (sponge) is used during force control. The bandwidths of DOB and RFOB are set to 500 rad/s., and $C_f = 5$.

Let us now consider how identification of plant parameters improves the stability of the robust force control system. Force control is implemented in the horizontal direction by using different nominal and identified mass values in the design of DOB and RFOB, respectively. The open loop gain is set to a fixed value by using $C_f \alpha = 2.5$. Fig. 11 shows the stability of the robust force control system. Fig 11a and 11b show that as the nominal mass of the plant is increased in the design of DOB, the stability of the robust force control system is improved. However, as shown in (5), the nominal mass cannot be increased freely due to the robustness constraint. Fig. 11c and 11d show that the value of identified mass that is used in the design of RFOB changes the stability of the robust force control system, significantly. A RFOB should be designed by using $\hat{M}_m \le M_m$, i.e., $\alpha \le \beta$ to improve the stability of the force control system. A hard environment (aluminum box) is used in the experiment. Since the transients between non-contact and contact motions are not treated, the wide impact forces are occurred in force control. It is obvious that the impact force can be suppressed by controlling the approaching velocity between non-contact and contact motions.

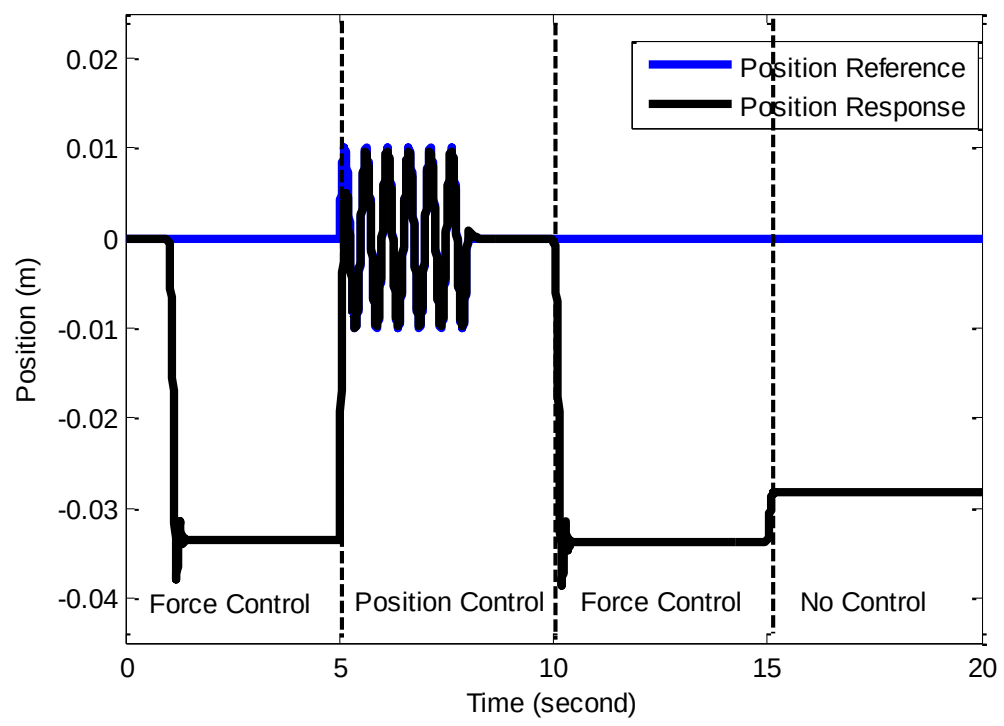

(a) Position control response

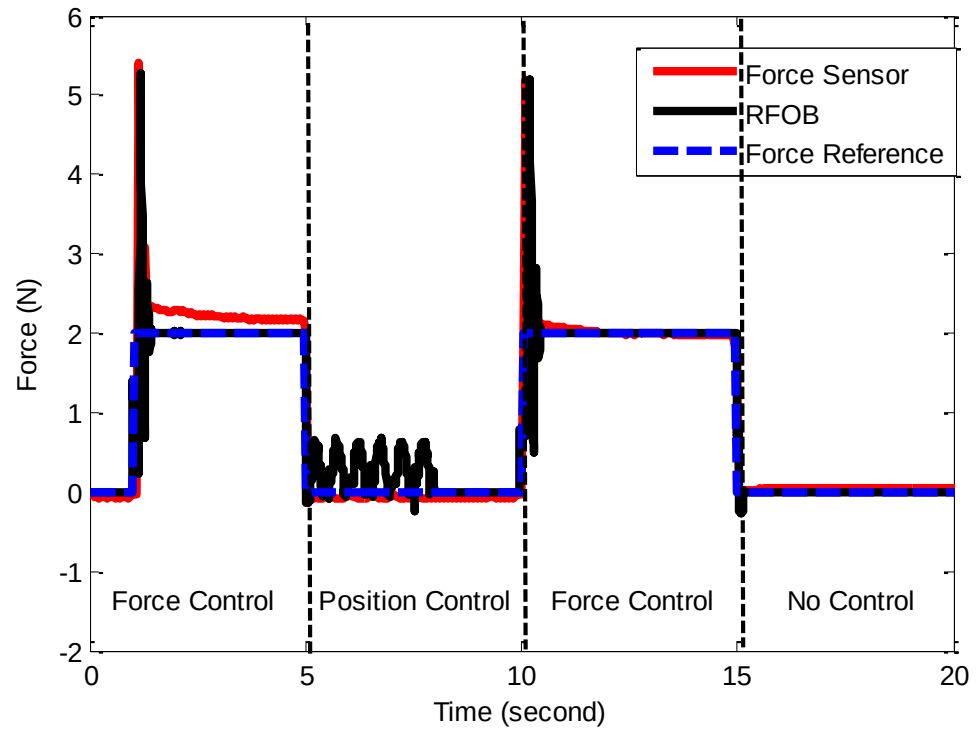

(b) Force control response

Fig. 10: Position and force control responses in vertical direction

So far, identification of environmental impedance has not been considered. Lastly, let us consider how identification of environmental impedance improves the robust force control system. The plant parameters are identified in free motion, and DOB and RFOB are designed by using $\alpha = 2$ and $\beta = 2$ to improve the stability. The force control response is shown in Fig. 11b when the adaptive algorithm is not implemented. The bandwidths of DOB and RFOB and the force control gain are tuned by using the adaptive algorithm with on-line and off-line parameter identification methods. Fig. 12 shows the force control responses when the adaptive algorithm is implemented. It is clear from Fig.11b and Fig.12 that the adaptive algorithm improves the force control response. However, as shown in Fig. 12a, the adaptive algorithm with on-line identification is affected by the dynamics of identification process during the transition between non-contact and contact motions, which is shown in Fig.5. In the

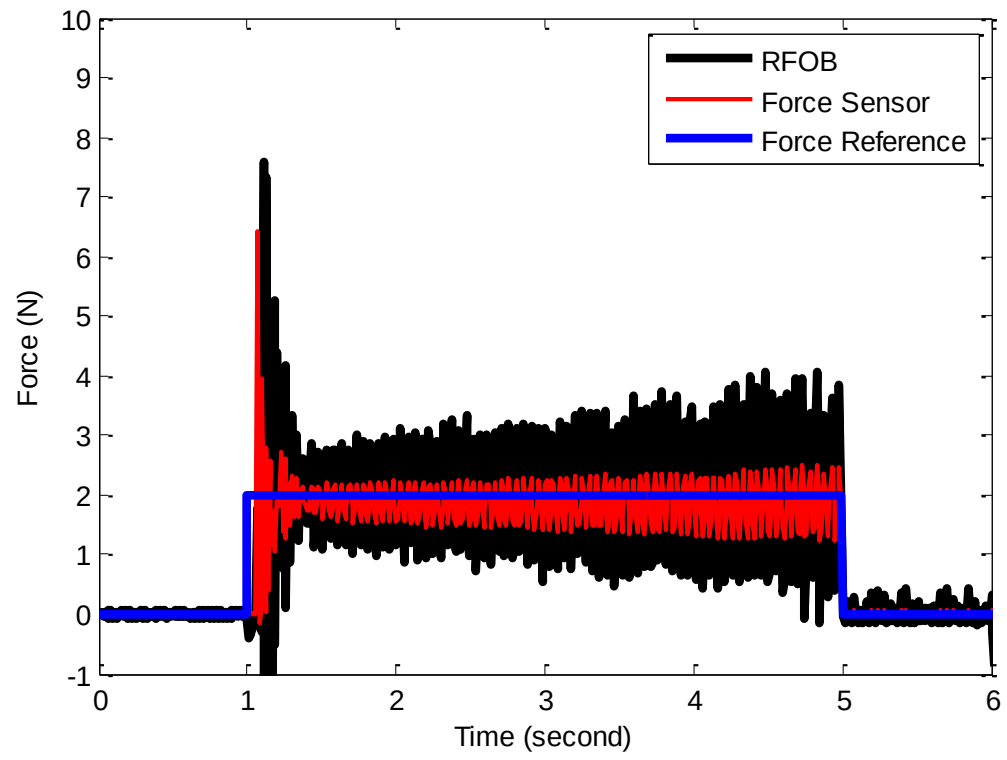

(a) $\alpha = \beta = 0.5, g_{DOB} = g_{RFOB} = 500\,rad/s$, and $C_f = 5$

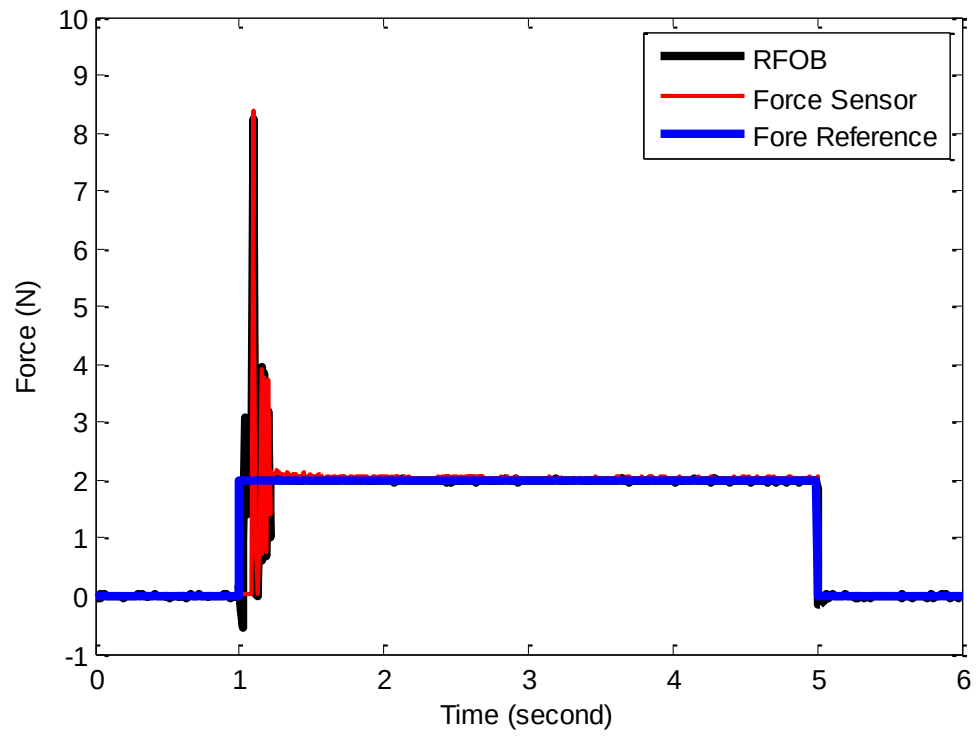

(b) $\alpha = \beta = 2, g_{DOB} = g_{RFOB} = 500\,rad/s$, and $C_f = 1.25$

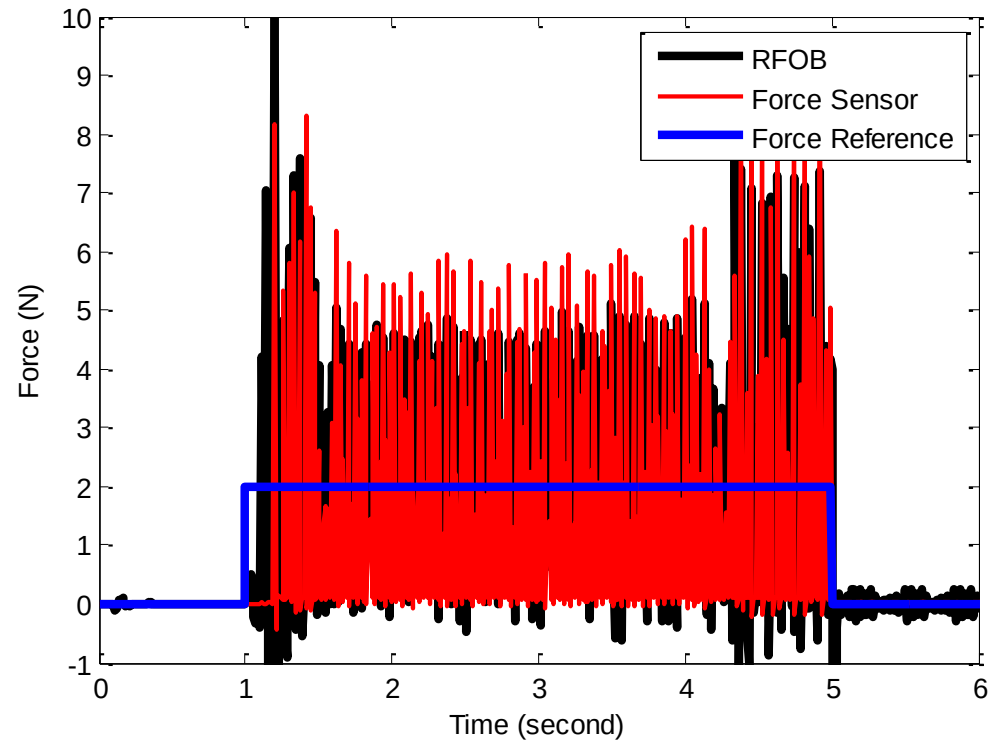

(c) $\alpha = 4, \beta = 2, g_{DOB} = g_{RFOB} = 500\,rad/s$, and $C_f = 0.625$

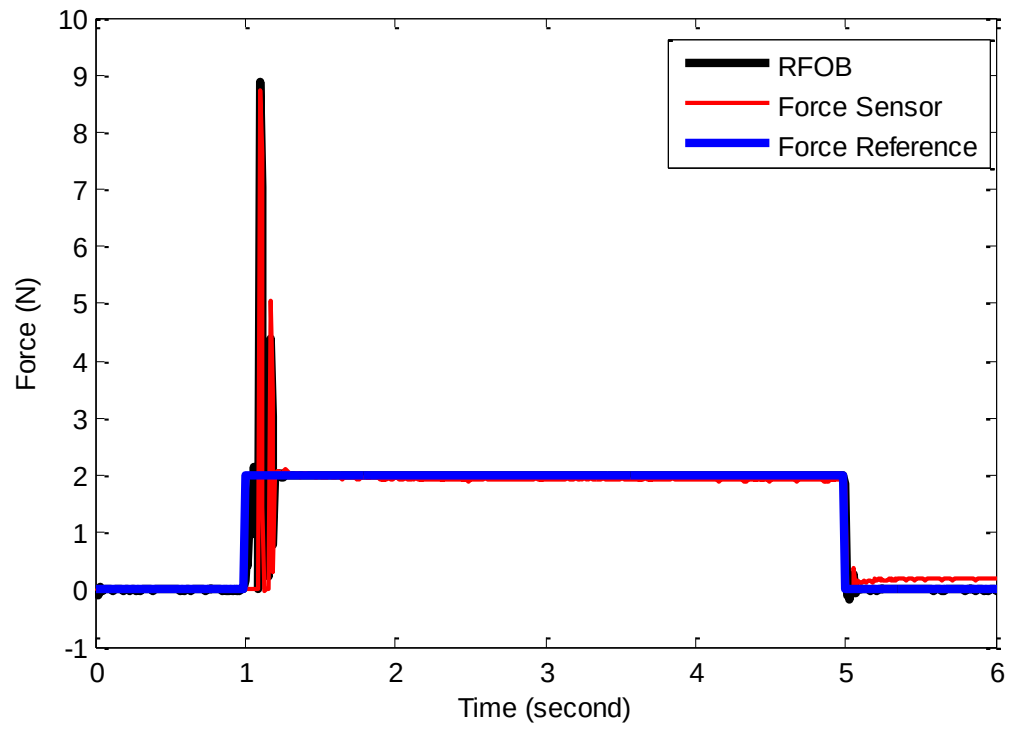

(d) $\alpha = 2, \beta = 4, g_{DOB} = 500\,rad/s, g_{RFOB} = 1000\,rad/s$, and $C_f = 1.25$

Fig. 11: Force control responses in horizontal direction

adaptive algorithm, the control parameters are not updated

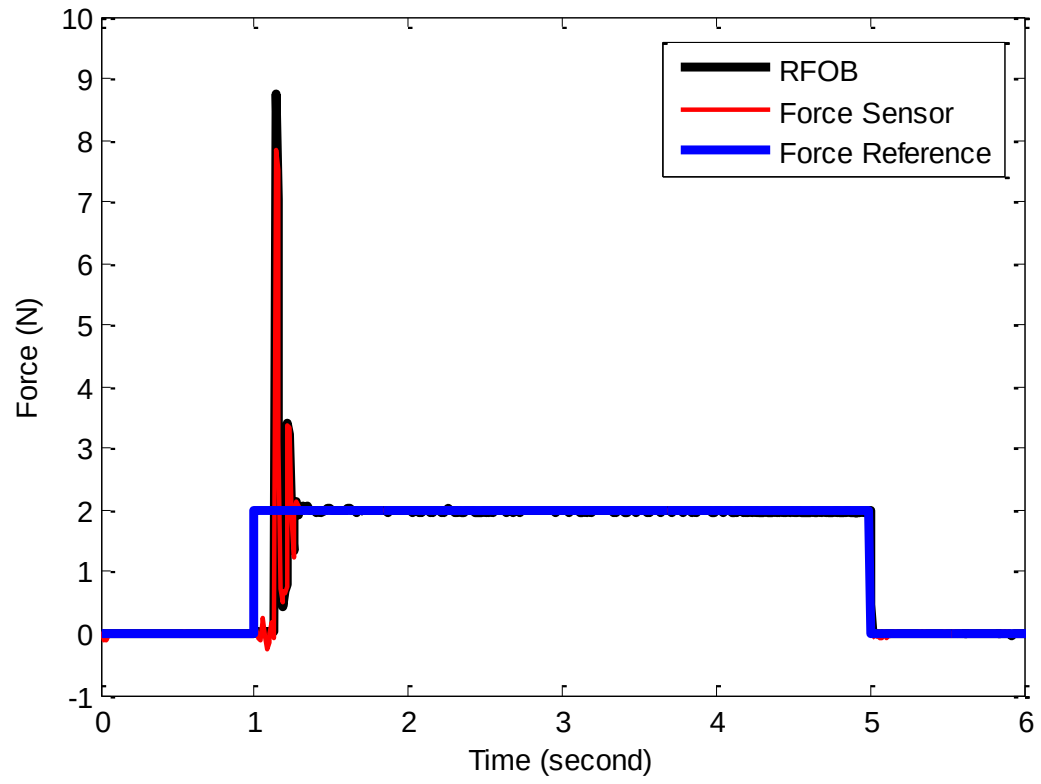

(a) Adaptive algorithm with on-line parameter estimation

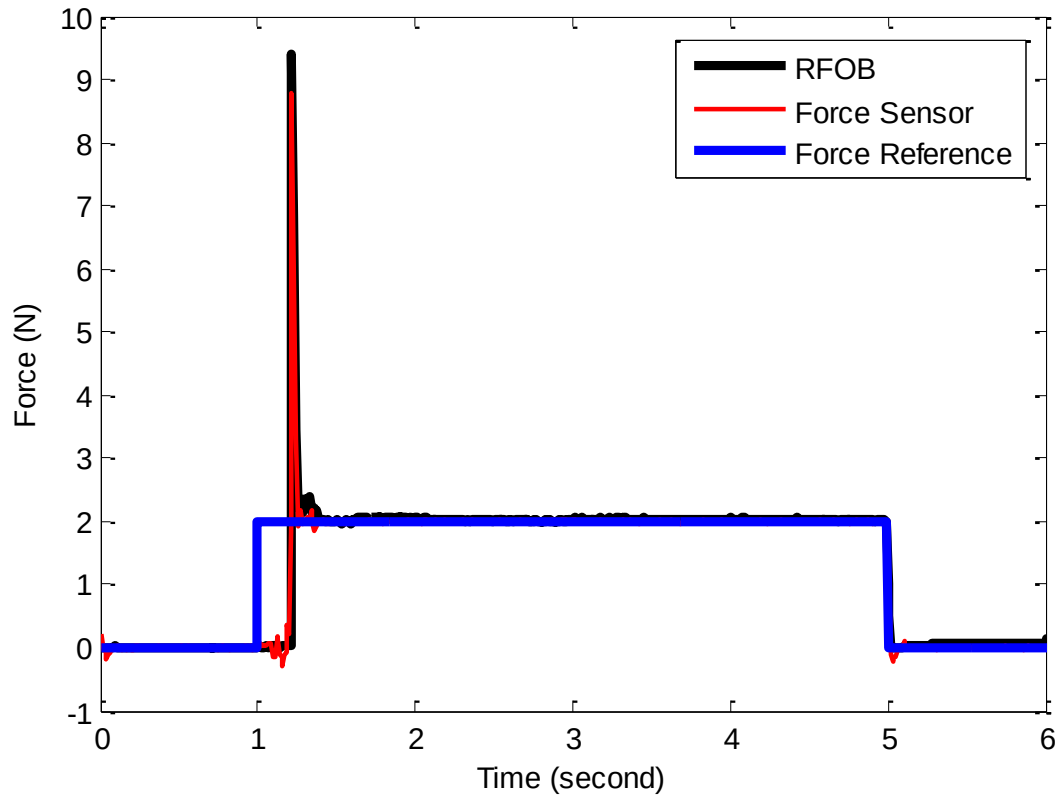

(b) Adaptive algorithm with off-line parameter estimation

Fig. 12: Force control responses in horizontal direction

during the transition, so oscillations cannot be suppressed precisely when on-line parameter identification is used. However, as shown in Fig. 12b, if environmental impedance is known a priori, which is impractical in many cases, then the oscillations can be suppressed precisely by using the proposed adaptive algorithm.

## CONCLUSION

This paper proposes a novel stability analysis and a new adaptive design methods for the RFOB based robust force control systems. It is shown that the stability of the robust force control system can be improved by increasing the nominal inertia in the design of DOB and the bandwidth of RFOB. The imperfect system identification in the design of RFOB may degrade not only the performance, but also the stability of the robust force control system significantly. Therefore, a new design constraint, that is $\alpha \leq \beta$, is proposed to improve the stability of the robust force control system. Against the conventional methods, which use actual inertias in the design of a DOB and a RFOB, this paper shows that not only the bandwidths but also the nominal and identified inertias of a DOB and a RFOB can be used as design parameters to improve the performance of force control. A new adaptive design method, which improves the stability and performance, is proposed by estimating plant parameters and environmental impedance. Simulation and experimental results verify the validity of the proposals.

## APPENDIX

*Solution of the Cubic Equations:*

*Property:* Let us consider a cubic polynomial and functions by using

$$\begin{aligned} & a_3x^3 + a_2x^2 + a_1x + a_0 = 0 \\ & \Delta = 18a_3a_2a_1a_0 - 4a_2^3a_0 + a_2^2a_1^2 - 4a_3a_1^3 = 27a_3^2a_0^2 \\ & \Delta_0 = a_2^2 - 3a_3a_1 \\ & \Delta_1 = 2a_2^3 - 9a_3a_2a_1 + 27a_3^2a_0 \end{aligned} \quad \text{(A1)}$$

Then,

- If $\Delta \geq 0$, the polynomial has three real roots.
- If $\Delta < 0$, the polynomial has imaginary roots.

and the roots of the polynomial are

$$\begin{aligned} x_1 &= -\frac{1}{3a_3}\left(a_2 + \Gamma + \frac{\Delta_0}{\Gamma}\right) \\ x_{2,3} &= -\frac{1}{3a_3}\left(a_2 + \frac{-1 \pm i\sqrt{3}}{2}\Gamma + \frac{\Delta_0}{\frac{-1 \pm i\sqrt{3}}{2}\Gamma}\right) \end{aligned} \quad \text{(A2)}$$

where $\Gamma = \sqrt[3]{\dfrac{\Delta_1 + \sqrt{\Delta_1^2 - 4\Delta_0^3}}{2}}$.

Let us first assume that $\xi^+ < 1$. Equation (35) should be solved by applying that $\eta = \eta^*$ where $\eta^* < 1$.

$$2\eta^*\xi^2\psi k^3 - \left(1 + 2\eta^*\xi^2\right)k^2 + 1 = 0 \quad \text{(A3)}$$

It can be easily checked that the polynomial has two positive real or imaginary and a negative real roots. To obtain a positive real $k$, all roots should be real. By using the *Property*, it can be shown that all roots are real if the following inequality is satisfied.

$$8\xi^6\eta^{*3} - \left(27\psi^2 - 12\right)\xi^4\eta^{*2} + 6\xi^2\eta^* + 1 \geq 0 \quad \text{(A4)}$$

By using the *Property*, it can be easily shown that (A4) is satisfied if the following conditions hold.

$$\begin{aligned} & \textit{if } \psi \leq 1, \textit{ then } \eta^* > 0 \\ & \textit{if } \psi > 1, \textit{then } 0 < \eta^* \leq \frac{\lambda_2}{\xi^2}, \textit{or } \eta^* \geq \frac{\lambda_3}{\xi^2} \end{aligned} \quad \text{(A5)}$$

where $\lambda_1 < \lambda_2 < \lambda_3$ are the roots of the polynomial given in (A4), in which $\lambda = \xi^2\eta^*$. Therefore, $\eta^*$ should be chosen small enough to satisfy (A5). Consequently, the roots of the polynomial given in (A3) can be calculated by using (A2).

Now, let us assume that $\xi^+ \geq 1$. Only the constraints on the bandwidth of a DOB should be considered, and $\eta$ can be chosen freely. Equation (35) can be rewritten as follows:

$$\frac{\frac{D_{env}}{M_m}}{\sqrt{\frac{K_{env}}{M_m}}} < \frac{4\xi^2\left(1-\psi k\right)k^2+1-k^2}{2\xi\left(1-\psi k\right)k} \le \frac{\frac{g_v}{2}+\frac{D_{env}}{M_m}}{\sqrt{\frac{K_{env}}{M_m}}} \qquad \text{(A6)}$$

or

$$\begin{aligned} &4\xi^2\psi k^3+\left(1-4\xi^2-2\xi\psi\vartheta\right)k^2+2\xi\vartheta k-1\ge 0 \\ &4\xi^2\psi k^3+\left(1-4\xi^2-2\xi\psi\delta\right)k^2+2\xi\delta k-1< 0 \end{aligned} \qquad \text{(A7)}$$

where $\vartheta = \frac{\frac{g_v}{2}+\frac{D_{env}}{M_m}}{\sqrt{\frac{K_{env}}{M_m}}}$ and $\delta = \frac{\frac{D_{env}}{M_m}}{\sqrt{\frac{K_{env}}{M_m}}}$.

It can be easily checked that the polynomials given in (A7) have one positive real and two negative real or imaginary roots when the coefficients of $k^2$ are positive and three positive roots when the coefficients of $k^2$ are negative. Therefore, the real positive solutions of the cubic polynomials can be obtained.

## References

[1] A. Mörtl, M. Lawitzky, A. Kucukyilmaz, M. Sezgin, C. Basdogan, S. Hirche. (2012, November). The Role of Roles: Physical Cooperation between Humans and Robots. *International Journal of Robotics Research*, vol.31, no.13, pp. 1656-1674

[2] N. V. Vasilyev, P. E. Dupont, P. J. del Nido. (2012, March). Robotics and imaging in congenital heart surgery, *Future Cardiology*. vol. 8, no. 2, pp. 285-296

[3] Qi Lu; Jianxun Liang; Bing Qiao; Ou Ma. (2013, February). A New Active Body Weight Support System Capable of Virtually Offloading Partial Body Mass. *Mechatronics, IEEE/ASME Transactions on* , vol.18, no.1, pp.11-20

[4] Ki-Young Kim; Ho-Seok Song; Jung-Wook Suh; Jung-Ju Lee. (2013, February). A Novel Surgical Manipulator with Workspace-Conversion Ability for Telesurgery. *Mechatronics, IEEE/ASME Transactions on*, vol.18, no.1, pp. 200-211

[5] A. Sabanovic, K. Ohnishi, ‘Part Two: Issues in Motion Control’, Advanced Motion Control, John Wiley & Sons, 2011

[6] M. Boudaoud, Y. Haddab, Y. Le Gorrec. (2013, June). Modeling and Optimal Force Control of a Nonlinear Electrostatic Microgripper, *Mechatronics, IEEE/ASME Transactions on*, vol.18, no.3, pp.1130-1139

[7] K. Ohnishi, M. Shibata, and T. Murakami. (1996, March). Motion control for advanced mechatronics. *Mechatronics, IEEE/ASME Transactions on*, vol. 1, no. 1, pp. 56–67.

[8] D. Tian, D. Yashiro, and K. Ohnishi. (2012, June). Wireless Haptic Communication under Varying Delay by Switching-Channel Bilateral Control with Energy Monitor. *Mechatronics, IEEE/ASME Transactions on*, vol. 17, no. 3, pp.488-498.

[9] Zi-Jiang Y., Fukushima, Y., Pan Q. (2012, September). Decentralized Adaptive Robust Control of Robot Manipulators Using Disturbance Observers", *Control Systems Tech. IEEE Trans. on*. vol. 20, no. 5.

[10] E. Sariyildiz, K. Ohnishi. (2013, May). Bandwidth constraints of disturbance observer in the presence of real parametric uncertainties, *European Journal of Control*. vol. 19. no. 3. pp. 199-205.

[11] E. Sariyildiz, K. Ohnishi. (2014) Stability and Robustness of Disturbance Observer Based Motion control Systems. *Industrial Electronics IEEE Trans. on (accepted)*

[12] H. Kobayashi, S. Katsura, K. Ohnishi. (2007, December). An analysis of parameter variations of disturbance observer for motion control,” *Industrial Electronics IEEE Trans. on.* vol. 54, no. 6, pp. 3413-3421.

[13] T. Murakami, F. Yu, and K. Ohnishi. (1993, April). Torque sensorless control in multi-degree-of-freedom manipulator. *Industrial Electronics IEEE Trans. on.* vol. 40, no. 2, pp. 259–265.

[14] Hongbing Li, K. Kawashima, K. Tadano, S. Ganguly, S. Nakano. (2013, February). Achieving Haptic Perception in Forceps’ Manipulator Using Pneumatic Artificial Muscle. *Mechatronics, IEEE/ASME Transactions on.* vol.18, no.1, pp.74-85, Feb. 2013

[15] S. Katsura, Y. Matsumoto, K. Ohnishi. (2006, June). Analysis and Experimental Validation of Force Bandwidth for Force Control, *Industrial Electronics IEEE Trans. on*, vol. 53, no. 3, pp. 922-928.

[16] S. Katsura, Y. Matsumoto, K. Ohnishi. (2007, February). Modeling of Force Sensing and Validation of Disturbance Observer for Force Control. *Industrial Electronics IEEE Trans. on* vol. 54. no. 1, pp. 530-538.

[17] T. Tsuji, T. Hashimoto, H. Kobayashi, M. Mizuochi, K. Ohnishi. (2009, February). A Wide-Range Velocity Measurement Method for Motion Control," *Industrial Electronics IEEE Trans. on* vol. 56, no. 2, pp. 510-519.

[18] H.W. Bode. Network Analysis and Feedback Amplifier Design. D. van Nostrand, NewYork, 1945.

[19] E. Sariyildiz, K. Ohnishi. (2014, March). A Guide to Design Disturbance Observer. *Asme Trans., Journ. of Dyn., Meas. and Cont*. vol. 136. no. 2. pp. 1-10 (021011)

[20] E. Sariyildiz, K. Ohnishi. (2013, October). Analysis the Robustness of Control Systems Based on Disturbance Observer. *International Journal of Control*. vol. 86, no. 10, pp. 1733-1743,

[21] C. Canudas de Wit, H. Olsson, K.J. Astrom, and P. Lischinsky. (1995, March). New model for control of systems with friction. *Automatic Control IEEE Transactions on*. vol. 40. no. 3. pp. 419–425.

[22] B. Yao. "Advanced motion control: From classical pid to nonlinear adaptive robust control", in Proc. 11th IEEE Inter. Workshop on AMC, 2010, pp. 815-829.

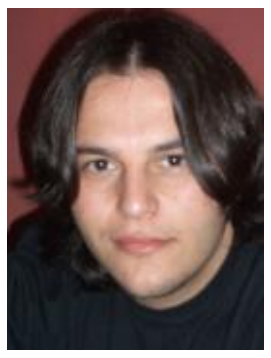

**Emre SARIYILDIZ** (S’11) received his Ms.C. degree in Mechatronics Engineering from Istanbul Technical University, Istanbul, Turkey in 2009. He is currently working toward the Ph.D. degree in integrated design engineering at Keio University, Yokohoma, Japan.

His main research interests are control theory and robotics.

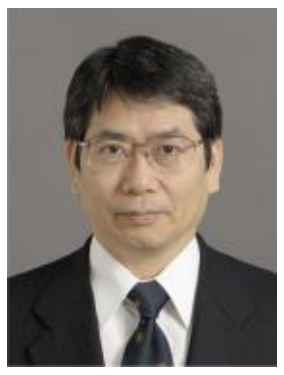

**Kouhei Ohnishi** (S’78–M’80–SM’00–F’01) received the B.E., M.E., and Ph.D. degrees in electrical engineering from the University of Tokyo, Tokyo, Japan, in 1975, 1977, and 1980, respectively.

Since 1980, he has been with Keio University, Yokohama, Japan. His research interests include mechatronics, motion control, robotics, and haptics.

Dr. Ohnishi received Best Paper Awards from the IEEJ and the Japan Society for Precision Engineering, and Outstanding Paper Awards at IECON’85, IECON’92, and IECON ’93. He also received the EPE-PEMC Council Award and the Dr.-Ing. Eugene Mittelmann Achievement Award from the IEEE Industrial Electronics Society in 2004.